# Liberalisation of the International Gateway and Internet Development in Zambia: The Genesis, Opportunities, Challenges, and Future Directions

Aaron Zimba
*Department of Computer Science & IT*
*Mulungushi University*
Kabwe, Zambia
gvsfif@gmail.com

Tozgani Fainess Mbale
*Department of Electrical & Electronics Engineering*
*University of Zambia*
Lusaka, Zambia
tozganimbale@gmail.com

Mumbi Chishimba
*Dept. of Information & Communications Technology*
*Zambia Revenue Authority (ZRA)*
Lusaka, Zambia
chishimba.mumbi@gmail.com

Matthews Chibuluma
*Department of IT & Information Systems*
*Copperbelt University*
Kitwe, Zambia
chibuluma.mathews@gmail.com

*Abstract*— Telecommunication reforms in Zambia and the subsequent liberalisation of the international gateway was perceived as one of the means of promoting social and economic growth in both the urban and rural areas of the country. The outcome of this undertaking propelled the rapid development of Internet which has evidently brought about unprecedented paradigm shifts in the use of Information and Communication Technologies (ICTs). It is indisputable that ICTs, and the Internet in particular, have revolutionalised the way we communicate today. Furthermore, the penetration of ICTs to other spheres of our daily lives is evidence enough that the impacts thereof go beyond mere communicative facets of our lives. However, many challenges arose in the implementation of telecommunications reforms. In order to achieve the status quo, government had to make strategic liberalisation policies in the telecoms sector that saw the opening up of the international communication gateways to the private sector. This is in tandem with the fact that the relationship between government (through its formulation of policies and regulations) and other stakeholders determines the ability of a country to generate and use advanced knowledge for industrial competitiveness. As such, in this paper, we present the genesis and evaluate the impacts associated with the telecommunications reforms and the subsequent liberalisation of international communication gateways, and Internet development in Zambia. We further consider the challenges this has brought about and discuss possible future directions. This is helpful in forecasting the future landscape of the ICT sector considering that the country seeks to achieve universal coverage of both Internet and communication facilities to all Zambians across the country.

## I. Introduction

On 22$^{nd}$ November 1994, Zambia became the first country in the entire sub-Sahara Africa to have full access to the Internet, apart from South Africa (Robinson, 1996). This placed the country as the fifth African country to have Internet connectivity. It's not a surprise that the first Internet connectivity was spearheaded by the University of Zambia and the government considering that the relationship between government and other stakeholders such as universities, determines the ability of a country to generate and use advanced knowledge for industrial competitiveness (Konde, 2004). The ongoing collaboration between government and other stakeholders has seen massive investments in the ICT sector.

The evolutionary development of Internet and the ICT sector cannot be fully understood without the consideration of the telecommunications (henceforth referred to as telecoms) industry. This is due to the fact that the telecoms industry has been known to be the backbone of Internet communications and various ICTs worldwide (Alden, 2008). From independence and up to 1997, the telecoms industry in Zambia had been dominated by the state owned company Zambian Post and Telecommunication Corporation (PTC) (Kaira, 2011). But just before then, in 1991 the new government had embarked on liberalisation ideologies in which it assumed the role of facilitator in most sectors of the economy. However, this was not entirely applicable to telecoms industry.

As such, the enactment of the Telecommunications Act ("Telecommunications Act," 1994) in 1994 led to the splitting up of PTC into two separate companies: the Zambia Postal Services Corporation (Zampost), and the Zambia Telecommunications Company (Zamtel). Zamtel, a product of liberalisation, remained the sole entity responsible for transmitting outbound and inbound international data traffic as well as being in control of the country's international gateway. This meant that all data traffic to the Internet had to go through Zamtel. This inherently gave Zamtel a monopolized competitive advantage. Although the international gateway had not yet been liberalized, it was in this same year that Zambia became the first sub-Sahara African country to have Internet connectivity after South Africa.

In 1994, Zamtel was the first Zambian telecoms company to provide mobile phone services. Traditionally, Zamtel (PTC) had been offering landline phone services. The year 1995 saw the entry of a private player (TELECEL, now MTN) into the telecoms industry. This entry was a result of the continued economic reforms commenced in 1991, the first since the famous "Mulungushi Reforms" of 1968 which had centred on nationalisation of various industries, the telecoms included. Airtel Zambia, formerly Zamcell (and later CELTEL), finally entered the Zambian market as a third mobile cellular service provider in 1997. To date, Zamtel remains the only mobile phone operator in Zambia which also offers Public Switched

Telecommunication Network (PSTN) services (ZICTA, 2015). After 2000, the tide in the telecoms industry started to shift towards mobile phones as opposed to landlines. This was a worldwide phenomenon as shown in Figure 1 (ITU, 2018). It is during this period after the introduction of private mobile network companies that the mobile userbase in Zambia began to grow. Henceforth, it's on this pretext that this paper focuses on the mobile phone sector of the telecoms industry.

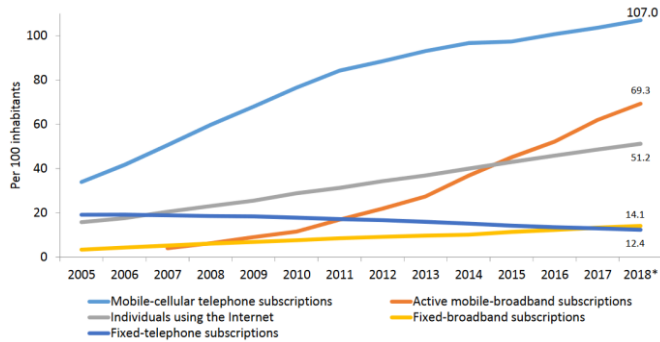

Figure 1. Global developments in mobile phone and ICTs usage

Notwithstanding the aforementioned worldwide strides in mobile phones adoption, the international gateway remained un-liberalised under the control of Zamtel. This lack of private sector players in the international gateway system led to very high international call and data tariffs, bottlenecks, as well as lack of investment in modern and more efficient technologies ("Zambia Competition Commission," 2008). Evidently, this was hindering the steady developments in communications and ICTs.

It was only after 2010 that the government sought to liberalise the international gateway system (M. Sumbwanyambe & Nel, 2011), a move that saw immediate participation of private sector players and subsequently an astounding ripple effect in reduced user prices for customers. The government had initially pegged the international gateway license fee at $12 million which was highly prohibitive and not reflective of the prevailing fees in the region. It was not until the reduction of the license fee from $12 million to $2 million, and finally to $350,000 that private players jumped unto the international gateway bandwagon. Since then, the telecoms sector with regards to Internet and the associated technologies have been developing rapidly. However, this has not come without a cost. There are many challenges that have ensued due to these developments.

As such, this paper discusses the genesis and impacts associated with liberalisation of Internet communication gateways and ICT developments in Zambia. We further consider the challenges this has brought about and possible future directions. Such an undertaking is useful in forecasting the future landscape of the ICT sector considering that the country seeks to achieve universal coverage of both internet and communication facilities to Zambians across the country (Mvunga, 2019).

## II. RELATED WORKS

Many studies have looked at the process and effects of liberalisation in Zambia. However, few of them have focused their efforts on the telecoms sector. (Kaira, 2011) looks at the state of competition in Zambia's telecommunications sector as at 2010. He argues that the benefits of liberalisation to the Zambian society have been immense, compared with the limited value offered by the closed fixed-line and international gateway markets. He acknowledges that there have been great strides to make sure that the Zambian ICT sector doesn't lag behind since the liberalisation policies of the 90's. However, he is quick to mention that reluctance towards competition policy in the international gateway market had limited the growth of the telecoms sector.

(Musonda, 2010) takes a critical analysis of the competition legal framework in Zambia in relation to the monopoly, and dominance by the state owned Zamtel Ltd of the international gateway in the telecoms industry. The author established the purported dominance by Zamtel of the international gateway by reviewing the competition legal framework where it was concluded that indeed Zamtel had had anti-competitive dominance over the international gateway. This had the effect of not stifling competition in the telecoms sector but the economy in general as well. It was further concluded that Zamtel's exclusive access to the international gateway was aided by other barriers to entry such as the $12 million exorbitant license fee for an international gateway service. The Competition and Fair Trading Act ("Competition and Fair Trading Act," 1994) did not appear to recognise single-firm monopoly behaviour under the prohibitions in section 7(2) which meant the monopoly and dominance by Zamtel was not provided for under the prohibitions. The author noted that this lacuna in the Act could easily argue that the conduct did not fall under the prohibited practices. The author further argued that the Zambia Competition Commission's (ZCC) lack of enforcement power exacerbates the firm's dominance. The dominance and unleveled playing field was attributed to the fact that Zamtel was a state-owned enterprise. Among other recommendations, the author recommended that the international gateway be fully liberalised and allow private sector players to acquire licenses by reducing the then exorbitant $12 million fee.

(Mambwe, 2015b) reviews the state of Internet technology in Zambia from a media perspective. He outlines a historical account of the birth of the internet in the country and proceeds to describe the current status of Internet technology in Zambia. He introduces the concept of Internet regulation in Zambia and points to ICT policies and Acts of parliament formulated by the government. In particular, the ICT policy () of 2007 is cited as the guiding document for ICT development in Zambia. The ICT sector is represented by a four-tier system that consists of policy-making, legal and regulatory framework, operators and consumers. Policy-making shapes the ICT industry and acts as a foundation for its development. The legal and regulatory framework includes institutions such as Zambia Information and Communications Technology Authority (ZICTA), Ministry of Justice, Ministry of Communications and so forth. Operators include telecoms service providers such as Zamtel, MTN, Airtel and so forth, while consumers include end-users, dealers in consumer electronics, consumer associations and corporate customers. ZICTA was formed by the 2009 ICT Act ("Information and Communications Technology Act," 2009) which repealed the 1994 Telecommunications Act and Radio Communications Acts ("Radio Communications Act," 1994) as an economic regulator with the authority to regulate tariffs for dominant players and agreements on interconnections. It's worth noting that even after these developments, the international gateway was not fully liberalised.

(Konde, 2004) looks at the Internet development in Zambia as a triple helix of government-university-partners. He identifies the relationship between government-university- (development) partners and outlines some important policy lessons where African universities could play a role as technology transfer agents. He underscores that underscores that technology transfer (the Internet inclusive) is not a simple mechanism of acquiring tools or means of production, processing or marketing, but involves innovation steps. he stresses that government participation in the development of new technologies is paramount. This is partly due to the fact that government supported the early inception of Internet at the university of Zambia despite not having had liberalised the international gateway which was the sole carrier of incoming and outgoing network traffic.

(Mambwe, 2014a) looks at the strides made after 20 years of Internet in Zambia and how that has impacted journalism among other things. He notes how the formulation of ICT policies and ICT-related Acts of Parliament have had tremendous effects in shaping the development of the Internet and ICTs. In particular, he points to the entrance of various ISPs, about 23, offering connection via Optic Fibre, VSAT, ADSL, Wi-fi Broadband, as well as mobile 3G, and 4G, and other technologies. He further noted the over tenfold increase in Internet subscriptions from 8,248 in 2001 to 92,642 at the end of 2012. However, increased access to the Internet brought its own challenges. He notes that abundant Internet access led to abuse and government raised several concerns over the seemingly reduced adherence to ethical guidelines particularly with the online media. He also notes the need for child online protection and safety as another emerging challenge due to the fact that Internet is easily accessible even by children.

III. LIBERALISATION OF THE ZAMBIAN INTERNATIONAL GATEWAY

An international gateway is defined as any facility that provides an interface to send and receive electronic communications (i.e., voice, data and multimedia images/video) traffic between a country's domestic network facilities and those in another countries (Alden, 2008). In the case of Zambia before the telecoms reforms, it was the nation's renowned Mwembeshi Earth Station (Pović et al., 2018; Tshibangu, 2017). Mwembeshi Earth Station was the backbone of all international communications where it transmitted outbound traffic from Zambia to the outside world and inbound traffic from the outside into Zambia's telecoms market. The Zambian telecoms market has three principal product markets; mobile telephone, fixed landline, and Internet. As such, Mwembeshi Earth Station was used to facilitate communication in all these product markets. The inbound and outbound telecoms signals or traffic that Mwembeshi administered can be summarised as; Internet traffic, telephone traffic (conversations), fax and telex messages, television and voice broadcast in and outside the country.

As such, being in control of Mwembeshi Earth Station gave PTC (Zamtel) the position of dominant player with a monopolistic competitive advantage. This inadvertently hindered the growth of the telecoms sector in Zambia. Even when the Zambian government liberalised the telecoms sector, the international gateway was not liberalised and it remained in the control of the state-owned telecoms enterprise Zamtel. After the introduction of private telecoms companies such as TELECEL and CELTEL, the Zambian government started facing pressure to fully liberalise the international gateway as was the trend in the region. There was a prevailing trend in the region of fully liberalizing the telecoms industry and ICT markets from the late 90's to 2010 as shown in Figure 1. It is clear from Figure 2 (Alden, 2008) that it's not only Zambia that hesitated to fully liberalise the international gateway despite having had liberalised other sectors of the telecoms industry.

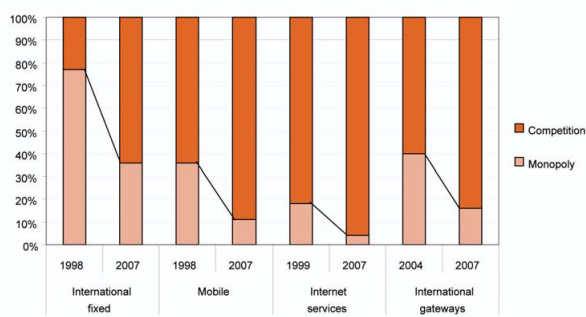

Figure 2. Trends in liberalization of international gateways and ICT markets, 1998-2007

In 2010, Zambia opened her international gateways to competition when the license fees were dropped to $350,000. This fee was reflective of what was prevailing in the region such as Uganda and Kenya were the international gateway license fees were $50,000 and $214,000 respectively. It must be noted that there was strong resistance to telecommunication reforms by interested groups such as political parties and civil societies including labour unions (M. Sumbwanyambe & Nel, 2011). However, the opening up of the international gateway saw the immediate participation of private players. Within a week of the reduction of international gateway license fees, private firms MTN and Zain (Airtel), entered the market and announced a reduction of international call rates by as much as 70% (Mfula, 2010). The international gateway market has been growing ever since. About five companies have rolled out fibre-optic networks with the largest being that of Fibrecom, a subsidiary of the Zambia Electricity Supply Corporation (ZESCO), whose network reaches all 10 provincial capitals and stretches to about 6 000 km. ZESCO's Fibrecom has direct international gateway connections to undersea cables through Tanzania, Malawi, Zimbabwe, Namibia and Botswana ("ITU Country Profile: Zambia," 2019).

By 2010, at the time of the full liberalisation of the international gateway, cumulative mobile phone subscribers for the three operators were only about 4,400,000 as shown in Table 1 (Habeenzu, 2010).

Table 1. Trends in Zambian Mobile Network Market 2010

| Service Provider | Total Subscribers | Market Share | Prepaid | Post Paid | Roaming | District Coverage |
|---|---|---|---|---|---|---|
| Cell Z | 152,581 | 3.5% | 152,015 | 566 | - | 44 |
| Zain Zambia | 3,089,270 | 70.1% | 3,076,377 | 12,893 | 3,089,270 | 72 |
| MTN Zambia | 1,164,831 | 26.4% | 1,159,657 | 5,174 | 1,164,831 | 72 |
| Sub-total | 4,406,682 | 100% | 4,388,049 | 18,633 | 3,254,101 | - |

After deregulation of the international gateways, the prices for both internet and voice access significantly fell and this resulted in increased mobile phone subscribers to more than 15 million by 2018 as depicted in Table 2 (ZICTA, 2019).

Table 2. Trends in Zambian Mobile Network Market 2018

| Year | | 2016 | 2017 | 2018 | Change |
|---|---|---|---|---|---|
| Airtel Zambia | Number | 4,971,355 | 5,332,496 | 5,897,968 | 565,472 |
| | Market Share | 41.4% | 39.7% | 38.1% | -1.6% |
| MTN Zambia | Number | 5,801,562 | 5,972,449 | 7,016,393 | 1,043,944 |
| | Market Share | 48.3% | 44.4% | 45.4% | 1.0% |
| ZAMTEL | Number | 1,244,117 | 2,133,594 | 2,555,909 | 422,315 |
| | Market Share | 10.4% | 15.9% | 16.5% | 0.6% |
| Total | | 12,017,034 | 13,438,539 | 15,470,270 | 2,031,731 |

Apart from increased mobile phone adoption and usage, liberalisation of the internet gateway resulted in increased internet access. This is supported by the fact that private mobile phone providers were also able to provide cheaper internet through their own gateways. It is important to note that intrinsic Internet Service Providers (ISP) were already operating on the market at that time.

The total domestic outgoing minutes on mobile networks increased by 40.2% by 2018 as shown in Table 3 (ZICTA, 2019). In the same manner, domestic incoming mobile voice traffic increased by 25.9%.

Table 3. Incoming and outgoing domestic mobile voice traffic in Zambia after telecoms liberalisation

| | 2016 | 2017 | 2018 | Percentage Change |
|---|---|---|---|---|
| Traffic domestic Incoming Minutes | 1,089,330,526 | 1,614,112,062 | 2,031,404,096 | 25.9% |
| Traffic domestic Outgoing Minutes | 11,500,328,930 | 9,967,432,124 | 13,975,890,482 | 40.2% |

On the contrary, compared to international incoming and outgoing voice traffic, there was a notable decline that continues to spiral downwards, at the time of writing this paper. With the liberalisation, and acquisition of international gateways by private players, one would expect international traffic to increase. However, this is not the case as shown in Table 4 (ZICTA, 2019). There was a significant decline of 39.5% in incoming international voice traffic and a 7.5% decline in outgoing international voice traffic by the end of 2018.

Table 4. Incoming and outgoing international mobile voice traffic in Zambia after telecoms liberalisation

| | 2016 | 2017 | 2018 | Percentage Change |
|---|---|---|---|---|
| Traffic International Incoming Minutes | 72,816,296 | 61,794,578 | 37,414,374 | -39.5% |
| Traffic International Outgoing Minutes | 86,757,866 | 43,146,291 | 39,891,483 | -7.5% |

The decline in outbound and inbound international traffic despite the liberalisation of the international gateway is mainly attributed to an increase in adopting internet-based applications like Skype, WhatsApp, and Viber among others to make international voice calls. In addition, adverse practices such as SIM boxing, a consequence of least cost routing could also explain this decline in international traffic despite liberalisation of the international gateway.

## IV. INTERNET DEVELOPMENT AFTER TELECOMS REFORMS

In the early 90s, Zambia was a notable pioneer of Internet in Sub-Sahara Africa. The first Internet was focused around the University of Zambia and the NGO community which finally materialised in 1994. The development of Internet during this period was boosted by liberalisation of the telecoms sector, most notably in 1996 when ZAMNET (Zambia premier ISP) was authorised to establish its own VSAT-based data gateway. Today, the Internet sub-sector is the most competitive in the telecommunications services industry in Zambia. This is because apart from traditional ISPs, the three telecoms companies provide Internet as well. Table 5 shows Internet penetration and usage in Zambia as at end of 2018 (ZICTA, 2019).

Table 5. Internet penetration and usage in Zambia

| Internet Usage | 2016 | 2017 | 2018 |
|---|---|---|---|
| Internet Subscription – Fixed Wireless | 35,919 | 36,121 | 44,711 |
| Internet users Per 100 Inhabitants | 0.22 | 0.2 | 0.3 |
| Mobile Internet Users - Smartphones/Blackberry/Dongles | 5,156,365 | 7,723,855 | 9,825,716 |
| Mobile Internet users Per 100 Inhabitants | 32.20% | 47% | 58.2% |
| Internet Users – fixed wireless & Mobile Internet Usage | 5,192,284 | 7,759,976 | 9,870,427 |
| Internet Usage Per 100 Inhabitants | 32.40% | 47.3% | 58.4% |

Mobile Internet users account for the largest type of Internet users. This is due to the fact that there are over 15 million mobile phone subscribers. In addition, the three mobile networks provide favourable Internet speeds.

The wide adoption of mobile Internet usage is vividly echoed by the access to Internet services by households across different types of technologies. Figure 3 (Sverige, ZICTA, 2019) shows that the main type of technology adopted by households as their main source of internet services is mobile broadband network via mobile phone. In 2018, this accounted for 78.4% of the total number of households that access internet services as shown in Figure 3.

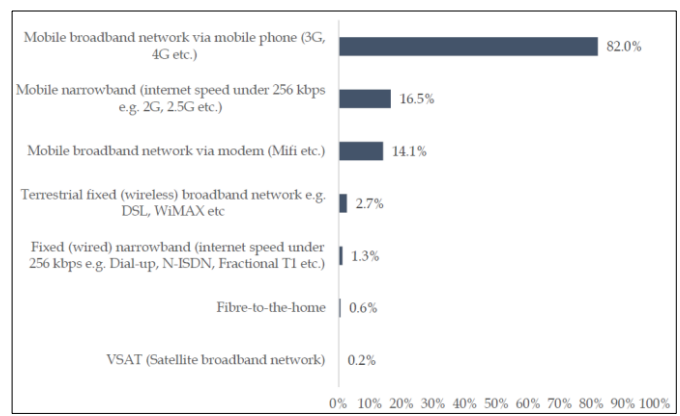

Figure 3. Access to Internet services by households

Mobile network operators now provide increased Internet speeds via 3G and 4G/LTE. Zamtel has rolled out 4.5G which has seen an increase in Internet usage (Zamtel, 2019). The Zambian government opened talks with Chinese tech giant Huawei on deploying 5G Internet in Zambia. Huawei Technologies and MTN mobile service provider launched a trial fifth Generation (5G) cellular network technology which provides even higher Internet speeds (Sinyangwe, 2019). The proportion of mobile network coverage by type of technology is shown in Figure 4 (Sverige, ZICTA, 2019).

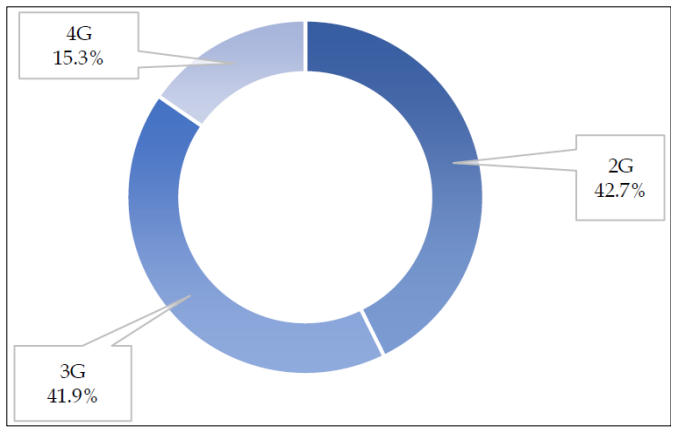

Figure 4. Mobile phone network coverage by type of technology

After the liberalisation of the telecoms sector and subsequent deregulation of the international gateway, access to Internet services increased to about 9.8 million active users by 2018 (ZICTA, 2019). This exponential increase in Internet usage could be attributed to improvements in price offerings on the market amid heightened competition among the vast service providers. There has been increased investment among providers after deregulation of the telecoms which has led to extensive coverage and the increased adoption of emerging technologies such as 4G/LTE (Newswire, 2018). The result has been further deduction in Internet access costs. The price of connecting to the Internet prior to the liberalisation of the international gateway was very high. For example, the monthly price of a 512 KBps home Internet connection from ZAMNET (the ISP pioneer in Zambia) in 2007 was about $80 while that of CopperNet was about $560 for a 32 KBps (Mulavu, 2007). These and other costings are shown in Table 6 and Table 7 respectively for the two aforesaid ISPs.

Table 6. The Cost of ZAMNET Broadband Services (Excl. VAT) in 2007

| Service description | Monthly subscription | Recommended no of users | Speed |
|---|---|---|---|
| Home user | ZMK326 000.00 | Single home user | 512 KBps |
| Single business user wireless Internet access | ZMK510 000.00 | Single corporate user | 512 KBps |
| Small office home user | ZMK817 000.00 | 2-4 users | 512 KBps |
| Small enterprise user | ZMK1 296 000.00 | 5-8 users | 512 KBps |
| Small to medium enterprise user | ZMK1 864 000.00 | 9-15 users | 512 KBps |

1US$ = ZMK4 132

Table 7. Cost of Broadband Services: CopperNet Solutions in 2007

| Name of Service | Installation Charge | Monthly subscription | Speed |
|---|---|---|---|
| Gold-32 ® | ZMK3 145 986.00 | ZMK2 327 654.43 | 32 KBps |
| Gold-64 ® | ZMK3 145 986.00 | ZMK4 481 781.00 | 64 KBps |
| Gold-128 ® | ZMK3 145 986.00 | ZMK8 077 978.00 | 128 KBps |
| Gold-256 ® | ZMK3 145 986.00 | ZMK12 062 814.00 | 256 KBps |
| VPV-32® | ZMK3 145 986.00 | ZMK1 789 116.50 | 32 KBps |
| VPV-64® | ZMK3 145 986.00 | ZMK1 866 204.50 | 64 KBps |
| VPV-128® | ZMK3 145 986.00 | ZMK5 092 063.50 | 128 KBps |

Source: CopperNET Solutions
1US$ = ZMK4,132

But after the deregulation of the international gateway and subsequent investments in the telecoms sector, Internet connection prices have continued to drop whereas the amount of available bandwidth has continued to rise. As such, the cost of Internet connectivity has continued to drop inversely proportional to the amount of available bandwidth. Figure 5 shows some of the latest pricing of Internet connection.

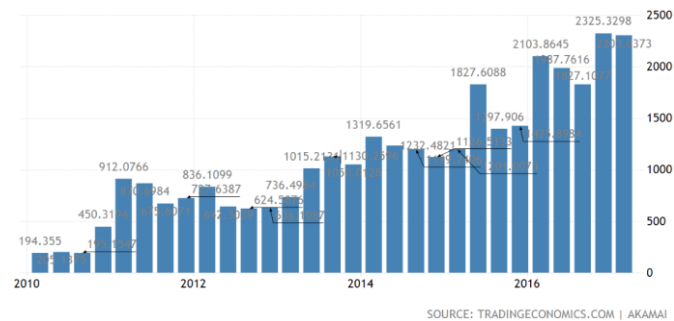

Figure 5. Cost of Zamtel Internet connection as at March 2020 ($1 = K15.80)

In comparison to the average cost of an Internet connection in 2007 ($100 - $500) with speeds of less than 512 KBps, the average monthly cost of a 3Mbps Internet connection as per Figure 5 is about $20. This is a clear indication that the amount of bandwidth available has continued to grow whereas as the cost thereof keeps on reducing.

Generally, the amount of Internet bandwidth available in Zambia since the liberalisation of the international gateways has continued to grow in a linear fashion as depicted in Figure 6 ("Zambia Internet Speed Data," 2020).

Figure 6. Zambia Internet Speed from 2010 - 2020

It's also worth noting that Zambia has been experiencing fluctuations in global Internet speeds as compared to other countries in the region (Akamai, 2017). The reasons behind these fluctuations, partly depicted in Figure 6, are beyond the scope of this paper.

V. OPPORTUNITIES

Telecoms reforms in Zambia have induced competitive pressure in the telecommunication market resulting in more employment opportunities. These opportunities are not only limited to the mobile network sector but to other ICT sectors like Internet Service providers, Carrier of Carriers, and Infrastructure Operators. Table 8 shows employment in the ICT sector by operator type 2016 – 2018 (ZICTA, 2019).

Table 8. Employment in the ICT sector: 2016 - 2018

|  | 2016 | 2017 | 2018 | Percentage Change |
|---|---|---|---|---|
| Mobile Network Operators | 1,417 | 1,308 | 1,217 | (7.0) |
| Internet Service Providers | 572 | 558 | 640 | 14.7 |
| Carrier of Carriers | 168 | 184 | 193 | 4.9 |
| Infrastructure Operator | 107 | 108 | 108 | 0.0 |
| Total | 2,264 | 2,158 | 2,050 | (5.0) |

As can be seen, the mobile network sector of the telecoms industry still leads the market in terms of employment opportunities. It is worth noting that this sector additionally interleaves with other sectors of society thereby indirectly providing additional jobs.

Deregulation of the telecoms industry has attracted massive investment in ICT infrastructure. As earlier noted, there are a number of service providers with multiple gateways which adds to the reliability of networks apart from the obvious advantage of not exhibiting a single point of failure. Huge investments in this industry has equally yielded higher output, faster network expansion and total factor productivity. Additionally, new wireless technologies such as 5G are seeing their way into the Zambian telecoms industry due to the ease of investing in the sector as a result of liberalisation. The mobile network service sector has also seen significant achievements in the area of cellphone rates reductions, improvement on mobile Internet 4G/LTE, infrastructure layout and the promotion of universal service in rural areas where ZICTA has been pioneering the installation of communication towers in rural areas (Mbuyu Sumbwanyambe, Nel, & Clarke, 2011). As such, all the three mobile service providers have reached almost every rural part of Zambia through the collaboration of ZICTA.

Full liberalisation of the mobile network sector of the telecoms industry has meant that new entrants bring tough competition which has often resulted in cheaper rates and improved quality of service apart from increased bandwidth. For example, Uzi's (Unitel International Holdings) attempt to enter Zambian market led to more than 70% fall in data prices (Mothobi, 2019). The Internet price reductions improved Zambia rankings in the Research ICT Africa Mobile Pricing Index from 31st in Q4 2017 to 13th place in Q1 2018, now being amongst the top 15 cheapest countries in Africa.

The development of the Internet in Zambia has seen a rise in a heavy online media. A number of online publications have emerged and traditional media stands at crossroads. Citizens no longer wait for the morning newspaper nor the evening news. As such, this has called for innovative opportunities of migrating online in order to have a persistent online presence. Most traditional media have online presence and this includes both radio and TV stations that stream live all and part of their broadcasts.

## VI. CHALLENGES

As earlier stated, ICTs have become a strong driving force in almost every aspect of development in society. However, in as much as ICTs are overwhelmingly powerful tools for development, they are paradoxically a "double-edged sword", providing many opportunities for individuals and organisations to develop but at the same time, bringing new challenges. Some of these challenges include cyber security, ethical regulation, user awareness and so forth.

As earlier noted, liberalisation of the telecoms sector, including the international gateway gave way to the abrupt rise of the Internet edge in Zambia. The Internet has brought its own challenges previously unheard of such as Internet crime. It often argued that laws governing internet crime in most countries especially with transitional economies seem to be limited. A case in point was when a suspected hacker defaced a government website by replacing the picture of the president with a cartoon. The suspected hacker was charged with defaming the head of state, but the case against him failed because there was no law in the country to deal with cybercrimes. As such, legislations have be proposed to address cybercrimes but there are concerns that such laws could be used to curb Internet access (Salifu, 2008).

The Internet provides various resources which social media. Social media in Zambia has become so widespread such that it is being abused by the citizens themselves. The two most common types of social media platforms in Zambia are Facebook and WhatsApp as shown in Figure 7 (Sverige, ZICTA, 2019).

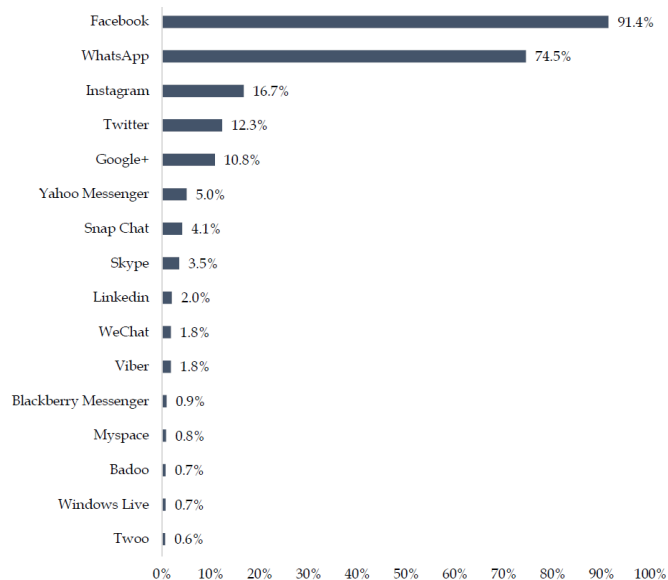

Figure 7. Usage of social media platforms; 2018

As such, it is not surprising that most of the social media abuse occurs on these two platforms. According to the ZICTA survey report (Sverige, ZICTA, 2019), the most prevalent known risks were associated with pornographic material, fake news and financial fraud accounting. Two of these three were associated with social media as shown in Figure 8.

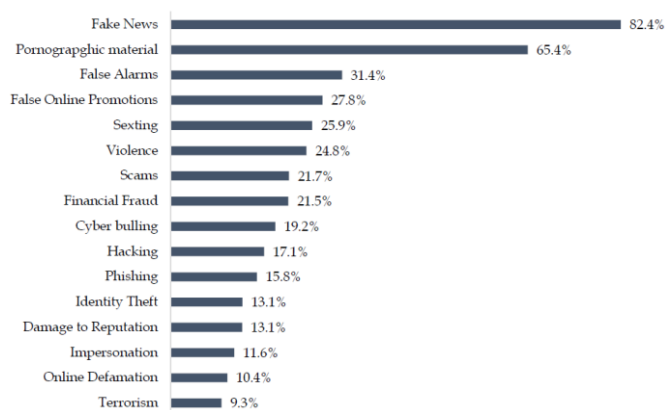

Figure 8. Incidence of risks while using social media, 2018

Apart from the above stated challenges, social media is rife with bullying. The type of bullying is not only directed towards adolescents but grown up as well. As such, the government has been on its toes to fight these vices which have proved to be challenging as some perpetrators might reside in other countries. In the same manner, there is a need to provide online child protection since children are being exposed to ICTs, and the Internet in particular, at a tender age.

New cyber security challenges that have gained prominace is cryptoviral attacks (A. Zimba & Wang, 2017; Aaron Zimba & Kunda, 2020) which is known to not only affect new Internet users but also experienced user who are oblivious of the paradigm shifts in attack structures of threats. In the case of ransomware attacks (Aaron, Mumbi, & Sipiwe, 2018; A. Zimba, Wang, & Chishimba, 2019), the attacker in such cases encrypts the victims files with resilient encryption without which it is impractical to recover the files until the demanded ransom is paid. The result has seen vicitms part away with a lot of money or indefinite loss of data due failure to pay the ransom.

Some of the other challenges encountered in this revolutionised industry include inadequate ICT policies to reflect what's currently obtaining, the absence of a strong legal framework encompassing all aspects of ICTs, lack of adequate cyber-attacks prevention and mitigation infrastructure, lack of relevant human capital in ICTs to meet the ever-evolving demand, and lack of sufficient sensitisation.

## VII. FUTURE DIRECTIONS

As the internet in Zambia clocks barely 3 decades, a lot has been achieved as evidenced in reduced Internet prices and increased bandwidth. but considering the fact that Zambia was the pioneer of Internet in Sub-Saharan Africa after South Africa, there remains a lot to be done. Likewise in the mobile network sector, there are challenges that need to be addressed such as security and quality of service. Generally, there are a number of other issues that the country needs to focus on relating to mobile network services and the internet. These are not limited to: the need to achieve universal access to ICTs for all as enshrined in the country's strategic plans, including the differently abled persons; improving on the country's cyber security infrastructure as the number of cyber-attacks have been increasing (Mambwe, 2015). Other issues that remain crucial in the future landscape of ICT development in Zambia include ensuring child online protection and improving user privacy and data protection. This requires rigorous legislation that ought to be supported by the very citizens it seeks to protect. Another issue of interest is addressing ethical concerns as a result of the available and easily accessible online platforms not limited to social media.

As more and more citizens are acquiring mobile phones and subscribing to cheaper Internet, various organisations such as banks, shops, schools, etc, are being challenged to provide their services using ICTs. Payment for service utilities such as water and electricity are already happening online. It is expected that more and more products and services will have an online alternative.

## VIII. CONCLUSIONS

This paper has presented an overview of the genesis, opportunities, challenges, and future directions associated with the liberalisation of the telecommunications industry in Zambia. Since the liberalisation of the telecoms industry, Zambia has seen great achievements in various sectors of ICTs. Liberalisation of the telecoms industry and the subsequent deregulation of the international gateway has fostered an exponential increase in mobile phone usage with increased internet bandwidth and reduced prices. It has been seen that the deregulation of the international gateway after liberalisation of the telecoms industry paved way for sustainable investments in ICTs, the most notable being the Internet. The result of such investments has been creation of new jobs and the emerging of new jobs associated with the telecoms industry both for the public and private sectors. And since technology is a "double-edged sword", developing countries with transitional economies such as Zambia need to understand the externalities that come with telecommunications reforms such privatization, liberalization and de-regulation. Cyber security concerns have emerged as challenges among other things. This likewise is a two-sided problem in that government has to deal with eventualities such fake news and abuse of social media among other things. The citizens, on the other hand, are concerned with online security, data privacy, government interference among others. As such, the onus in on both the government and its citizens to work in collaboration and harness the advantages that come with ICT developments in the telecoms sector as well as address the arising challenges amicably.